\documentclass[superscriptaddress,twocolumn,prb]{revtex4-1}
\usepackage{graphicx}
\usepackage{dcolumn}
\usepackage{multirow}
\usepackage{booktabs}
\usepackage{bm,color}
\usepackage{braket}
\usepackage{amsmath,amssymb}
\usepackage[colorlinks,linkcolor=blue,hyperindex,CJKbookmarks]{hyperref}
\usepackage{epstopdf}

\newcommand{\blue}[1]{\textcolor{blue}{#1}}
\newcommand{\mg}[1]{\textcolor{black}{#1}}

\newcommand\ba{\begin{eqnarray}}
\newcommand\ea{\end{eqnarray}}
\def\be{\begin {equation}}
\def\ee{\end {equation}}
\def\ber{\begin {eqnarray}}
\def\eer{\end {eqnarray}}
\def\bers{\begin {eqnarray*}}
\def\eers{\end {eqnarray*}}

\begin {document}

\title{\bf Effect of chirality imbalance on Hall transport of PrRhC$_2$}

\author{Banasree Sadhukhan}
\email{banasree@kth.se}
\affiliation{ KTH Royal Institute of Technology, AlbaNova University Center, 10691 Stockholm, Sweden}
\affiliation{Leibniz Institute for Solid State and Materials Research (IFW) Dresden, Helmholtzstr. 20, 01069 Dresden, Germany}

\author{Tanay Nag}
\email{tanay.nag@physics.uu.se}
\affiliation{Department of Physics and Astronomy, Uppsala University, Box 516, 75120 Uppsala, Sweden}
\affiliation{Institut f\"ur Theorie der Statistischen Physik, RWTH Aachen University, 52056 Aachen, Germany}

\begin{abstract}
Much has been learned about the topological transport in real materials. 
We investigate the interplay between magnetism and topology in the magneto-transport of PrRhC$_2$. The four-fold degeneracy reduces to two-fold followed by non-degenerate Weyl nodes when the  orientation of the magnetic quantization axis is changed from easy axis to body-diagonal through face-diagonal. This engenders chirality imbalance between positive and negative chirality  {\mg {Weyl nodes}} around the Fermi energy. We observe  a significant enhancement in the chiral anomaly mediated response such as  planar Hall conductivity and longitudinal magneto-conductivity, due to the emergence of  chirality imbalance upon orienting the {\mg {magnetic quantization axis}} to body-diagonal. \textcolor{black}{The angular variations of  the above quantities for different {\mg {magnetic quantization axis}} clearly refer to the typical signature of {\mg {planar Hall effect}} in  {\mg {Weyl semimetals}.}}   We further investigate the profiles of anomalous Hall conductivities as a function of Fermi energy to explore the effects of symmetries as well as chirality imbalance on Berry curvature.

\end{abstract}

\maketitle

{\it \blue {Introduction:}} The gapless topological   systems, for example, Dirac and Weyl semimetal (WSM), have received huge attention in recent times in the context of solid state research.    The 
four-fold degenerate Dirac points split into two two-fold degenerate Weyl nodes (WNs) once the time reversal  symmetry and  inversion  symmetry are broken separately or simultaneously \cite{vazifeh, cortijo}.
The WNs with opposite chiralities, 
designated by topological charge namely Chern number, act like a monopole and an antimonopole of Berry flux in momentum space.  {\mg{The inversion symmetry}} is broken in the transition-metal monopnictides (TaAs-family), dichalcogenides (MoTe$_2$-family) \cite{Lv_2015, Huang_2015, Hasan_2015, Wu_2016, Jiang_2017} while  magnetic WSMs (Co$_3$Sn$_2$S$_2$, Heusler alloy-family, rare earth carbides-family) \cite{enke,  PhysRevResearch.1.032044, yang,  sadhukhan} break {\mg {time reversal  symmetry}}.
\textcolor{black}{Upon increasing the tilt strength of the conical dispersion, the point-like Fermi surface of a type I WSM at the WN energy
can acquire a pocket-like shape resulting in a type-II WSM \cite{VOLOVIK2014514,YXu15,Soluyanov2015}.}

The WSMs exhibit several chirality related transport under the application of external magnetic  and electric fields. The non-conservation of chiral Weyl fermions at two WNs, referred  to as the chiral anomaly \cite{Adler:1969, Bell:1969, Nielsen:1981, Nielsen:1983}, \textcolor{black}{results in} a hallmark signature of negative longitudinal magnetoresistance for WSM under parallel magnetic and electric fields \cite{Zyuzin:2012,PhysRevB.86.115133,Wan_2011, Xu:2011,Goswami:2013}. The co-planar arrangement of electric and magnetic fields can lead to
planar Hall effect (PHE) \cite{Nandy_2017,Nandy_2018,Tanay_2018} that is  fundamentally different from the Lorentz force driven conventional Hall
effect.  Importantly,  
chiral anomaly mediated negative  {\mg {longitudinal magnetoresistance}} and PHE in Dirac semimetals and WSMs have been experimentally observed  \cite{Huang_2015,Jia_2016,Xu_2016,Erfu_2016,Li_2018,Liang_2018,Wang_2018,Chen_2018,Kumar_2018,Singha_2018,PhysRevB.98.081103,PhysRevB.98.121108,  PhysRevB.98.161110}.  {\mg {A giant PHE has been theoretically predicted in topological materials due to chiral anomaly \cite{PhysRevB.96.041110}}}. Interestingly, in the absence of electric field, charge transport occurs due to {\mg {chiral magnetic effect}} in WSMs with non-degenerate WNs \cite{PhysRevLett.109.181602, PhysRevB.88.104412, Zhou_2013,  PhysRevLett.116.077201,  PhysRevB.92.085138,PhysRevLett.115.117403,li2016chiral}. The anomalous Hall effect (AHE) is, on the other hand, a key signature of non-trivial Berry curvature  of magnetic WSM in the absence of magnetic field \cite{PhysRevLett.113.187202,Shekhar9140,zyuzin2016intrinsic}. \textcolor{black}{Notice that the Berry curvature in {\mg {inversion  symmetry}} broken family of non-magnetic WSMs is also instrumental to yield the non-linear transport properties extending the quantum topological transport beyond the realm of linear regime  \cite{de2017quantized,PhysRevB.104.245122,PhysRevB.103.144308,PhysRevB.103.245119,nag2022distinct,Das21}. }

In a very recent study, 
the \textcolor{black}{chirality imbalance}
has been engineered near the Fermi energy upon suitably tuning the magnetic quantization axis (MQA) in {\mg {time reversal  symmetry}} and {\mg {inversion  symmetry}} broken RMC$_2$-family (R=rare earth, M=transition metal) of WSMs \cite{sadhukhan}. 
To this end, considering  one of the candidate materials from the above family namely \textcolor{black}{PrRhC$_2$}, we seek answer to the following question: What is the consequence of chirality imbalance, caused by the canting of MQA, in the PHE? Notice that PHE has been extensively studied in the context of model Hamiltonians of WSMs while it has been unexplored so far from the perspective of real materials. Therefore, our study can become useful to predict experimental signatures \textcolor{black}{regarding} the interplay between magnetism and topology in  the context of PHE.

\begin{figure}[ht]
\centering
\includegraphics[width=0.5\textwidth,angle=0]{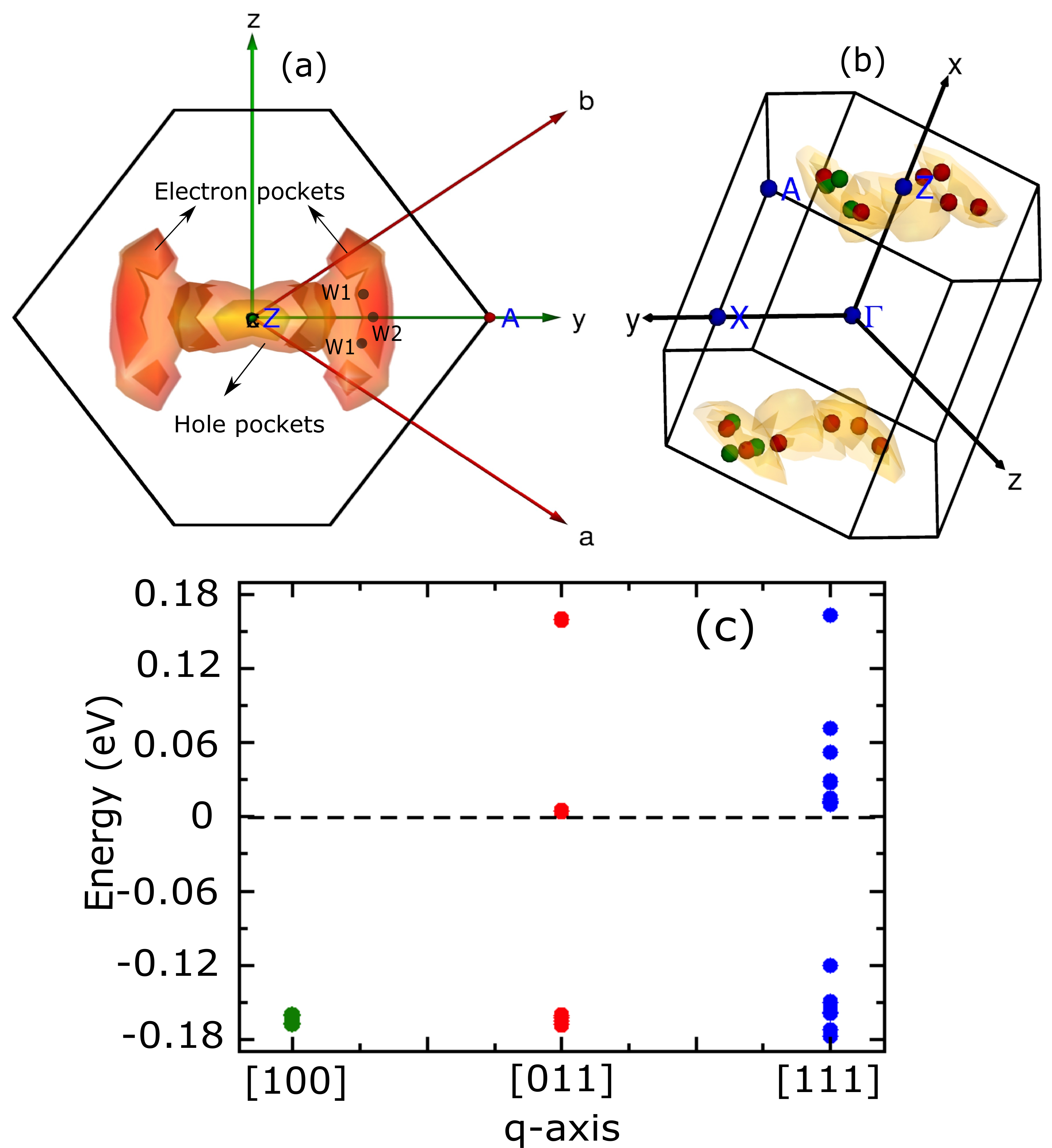}
\caption{(a) The top view of bulk three-dimensional (3D) Fermi surface  with the WNs in the BZ for MQA $[100]$. (b) The bulk 3D Fermi surface with the positions of the WNs in the BZ.  The green and red points represent the WNs for MQA $[100]$ and $[011]$, respectively. Changing the MQA from $[100]$ to $[011]$, the WNs split in energy and scattered in the BZ along $Z-A$ direction. The pockets appear at the zone boundary.  (c) Energy of the WNs for MQA  $[100]$, $[011]$ and $[111]$ configuration within the energy window $(-0.19, 0.19)$ eV.}
\label{fig:stuc_wp} 
\end{figure}

\begin{figure}[ht]
\includegraphics[width=0.51\textwidth,angle=0]{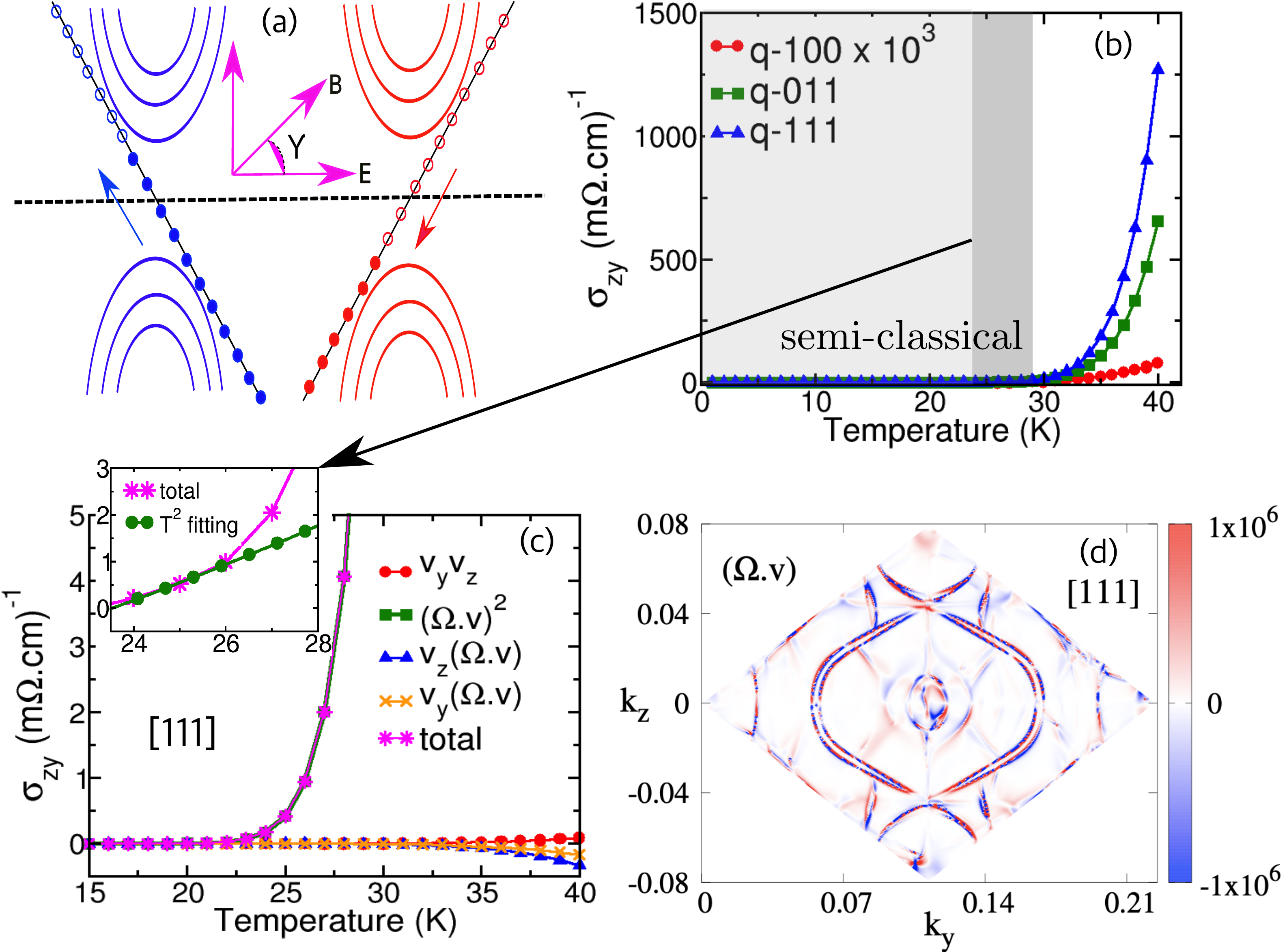}
\caption{(a) Left (blue) WN  is over populated as compared to the  right (red) WN due to chiral anomaly in presence of co-planer $\mathbf{E}$ and $\mathbf{B}$ field at an angle $\gamma$.
(b) The temperature profile of PHC for different orientations of MQA where
light-grey, grey and white regions respectively exhibit substantially low, moderate and large values of PHC.   The magnitude of PHC for ${\bf q}$-$100$ is significantly less as compared to other ${\bf q}$'s.   (c) The term-wise breakdown of PHC for $\bf{q}$-$111$ suggests that the chiral anomaly term, containing $(\mathbf{\Omega_{k}}\cdot\mathbf{v_{k}})^2$ factor, contributes maximally. The grey region in (b) is depicted as the inset where the PHC  varies quadratically with temperature between $22~ {\rm K}<T < 26 ~{\rm K}$. (d) The quantity  $(\Omega \cdot v) \equiv \sum \int dk_x(\mathbf{\Omega_{k}}\cdot\mathbf{v_{k}})$ in $k_y-k_z$ plane  for $\bf{q}$-$111$. 
We consider $B=9$ T,  $\gamma=\pi/3$, $\mu=10$ meV. 
}
	\label{fig:phe_temp} 
\end{figure}

\begin{figure}[ht]
\includegraphics[width=0.5\textwidth,angle=0]{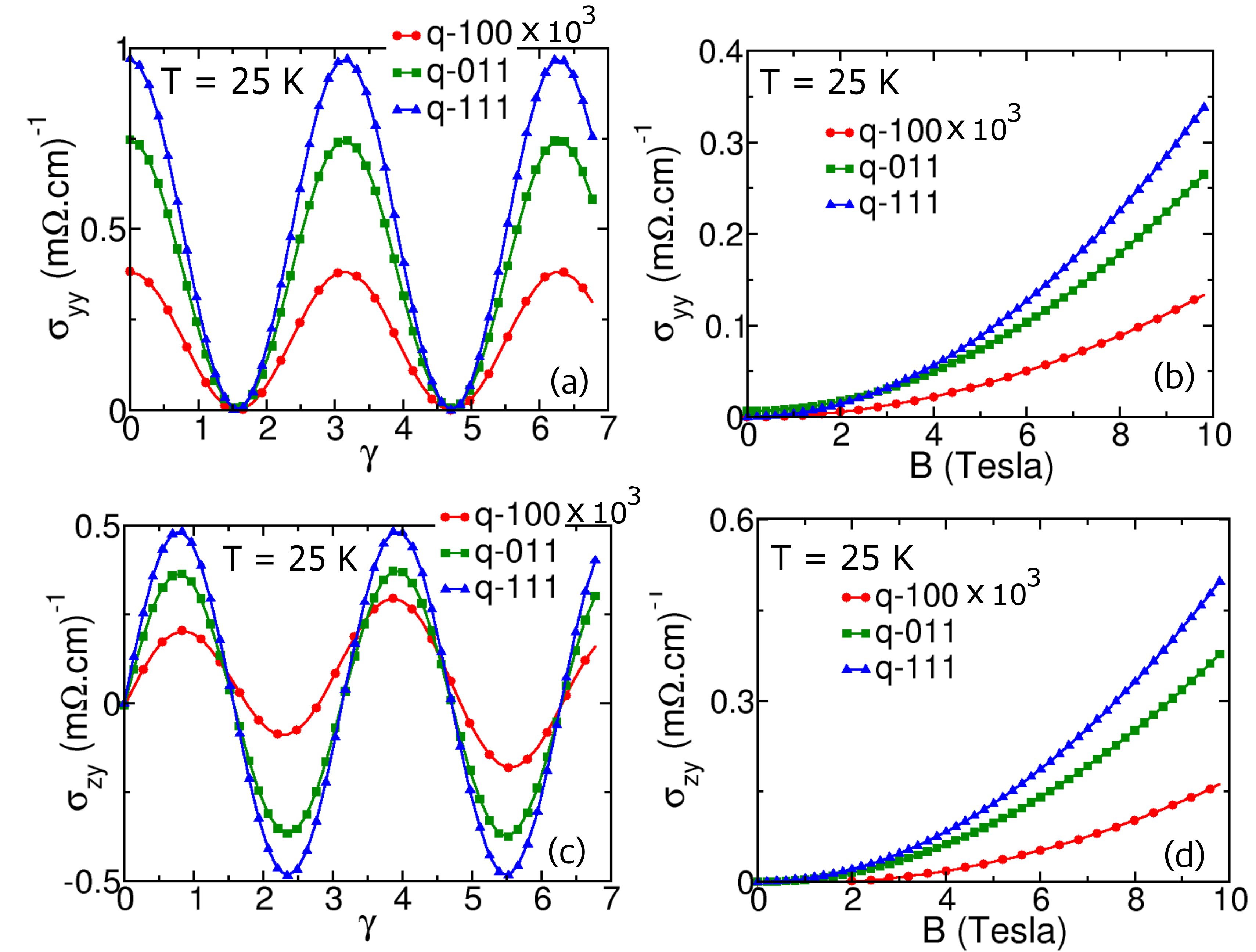}
\caption{Plots (a) and (c) show angular profile of  LMC $\sigma_{yy} \propto \cos^2 \gamma$ and PHC $\sigma_{zy}$ nearly $\propto \sin \gamma \cos \gamma$    at $T=25$ K, and $B=9$ T upon  changing MQA.   We repeat (a) and (c) as a function of $B$ for fixed $\gamma=\pi/3$ in (b) and (d), respectively, where $B^2$-dependence is clearly noticed. We consider $\mu=10$ meV.
}
	\label{fig:phe} 
\end{figure}

\begin{figure*}[ht]
\centering
\includegraphics[width=0.9\textwidth,angle=0]{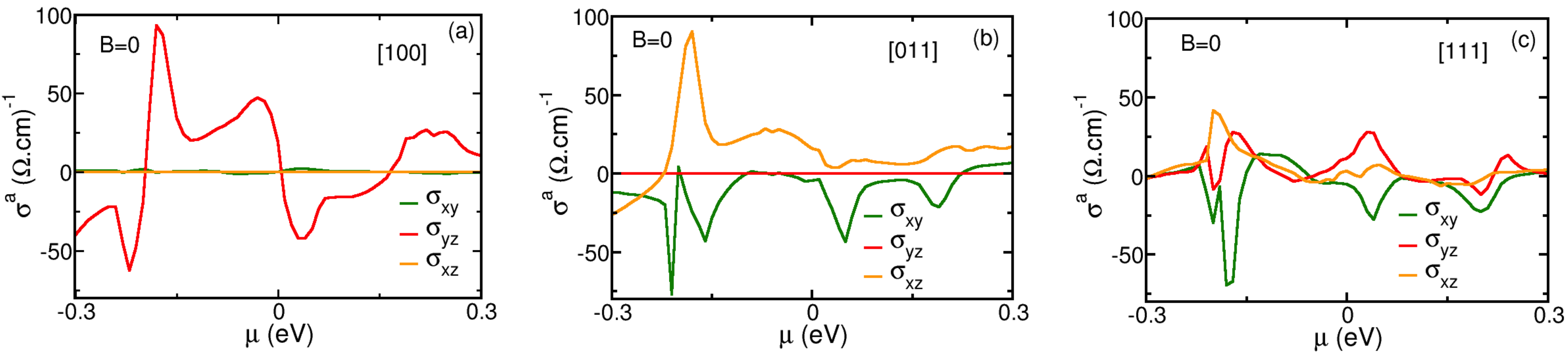}
\caption{The AHCs as the function of Fermi energy $\mu$ for MQA (a) $[001]$, (b) $[011]$, and (c) $[111]$. We consider $T=0$ K.  {\mg{When the MQA is varied from $[001]$ to $[111]$, the  symmetry of the system is systematically reduced. As a result, only one  component of AHC $\sigma_{yz}$ is substantially strong in (a). Two components of AHC  $\sigma_{xy}$ and $\sigma_{xz}$ contribute for $[011]$ in (b). For $[111]$ in (c),  all components of AHC  contribute.}}}
	\label{fig:ahe} 
\end{figure*}

In this paper, starting from the ferromagnetic  state with internal MQA  $\bf{q}$-$100$ where WNs are four-fold degenerate, we show that the degree of degeneracy can be lowered upon canting the MQA towards the body-diagonal (see Fig. \ref{fig:stuc_wp}). This is caused by the reduction in the combination of spatial point group and temporal symmetries of the material. Next, considering the setup of PHE,  we investigate the interplay between magnetism and topology in LM conductivity (LMC) and PH conductivity (PHC) [see Fig. \ref{fig:phe_temp}].
We remarkably find that the PH transport coefficients are maximally governed by the chiral anomaly while the breaking of the degeneracy of WNs, causing chirality imbalance, plays the key role in augmenting their amplitudes (see Fig. \ref{fig:phe}). \textcolor{black}{The angular dependencies we find for the above quantities by varying MQA are considered to be hallmark signature of PHE in WSMs.} Finally, we investigate AH conductivity (AHC) in absence of any external fields to highlight the role of the above symmetries on Berry curvature for different MQA (see Fig. \ref{fig:ahe}).

{\it \blue {Effect of {{magnetic quantization axis}} on the degeneracy of {{Weyl nodes}}:}}  
We use the framework of Full-potential local-orbital minimum-basis (FPLO)\cite{fplo_web, klaus1999, pbe1996}, based on  density functional theory (DFT). We find that PrRhC$_2$ is  a type-II WSM where  the electron and hole pockets simultaneously exist at WNs energies (see Fig. \ref{fig:stuc_wp} (a)) \cite{sadhukhan, sm_text}. Based on a hypothetical non-magnetic  PrRhC$_2$, one can expect to have  the full magnetic symmetry: $G = S + S {\cal T},$ where, $S = \{E, m(x),m(y),C_2(z)\}$ is the space group of crystal symmetry ($E$ is the identity operator) and ${\cal T}$ denotes  the {\mg {time reversal  symmetry}}. Notice that the evolution of WNs in   Brillouin zone (BZ) and \textcolor{black}{band structure have} been studied by rotating the MQA $\bf q$ \cite{PhysRevResearch.1.032044}. \textcolor{black}{The MQA represents the spin quantisation
axis; this can be oriented in a different direction for the crystal lattice when the spin-orbit coupling (SOC) is included in DFT calculations. The rotation in MQA yields
different ground-state, associated with tight-binding Wannier Hamiltonian, having different crystal symmetries. Such a tuning of MQA has already been theoretically investigated \cite{Zhisheng18,Laurent19,Xindong96} as well as experimentally demonstrated by applying external magnetic field \cite{chikazumi1997physics}, temperature \cite{drijver1976magnetic,albertini2004magnetocrystalline}, hydrostatic pressure \cite{Zhisheng18}, and uniaxial strain \cite{lyubina2005magnetocrystalline}.}


\par \textcolor{black}{We consider  PrRhC$_2$, hosting itinerant electrons, due to its magnetic property for partially occupied Pr-f shell  \cite{Steinbeck01,steinbeck2001magnetocrystalline}. The SOC tends to align Pr-f shell with the spin quantization axis and  the crystal field aligns it according to the crystal lattice. This results in magnetic anisotropy and  relativistic
symmetry reduction; we intend to study their effect on transport within ab-initio treatment.} For $\bf{q}$-$100$, $\bf{q}$-$011$ and $\bf{q}$-$111$, 
the following symmetries are respected $\{E,S_x, S_{y,z}{\cal T}\}$, $\{E,S_{x}{ \cal T}\}$, and $\{E\}$, respectively, while $\{S_{y,z},S_{x}{\cal T}\}$, $\{S_{x,y,z},S_{y,z} {\cal T}\}$ and $\{S_{x,y,z},S_{x,y,z}{\cal T}\}$ symmetries are absent (See Table \ref{tab:aniso}). These allow the WSM to host $4$-fold degenerate [$2$-fold degenerate]  WNs for $\bf{q}$-$100$ [$\bf{q}$-$011$]  at ($\pm k_x,k_y,\pm k_z$) [($k_x,k_y,k_z$) and ($k_x,-k_y,-k_z$)]  in the BZ. While for $\bf{q}$-$111$, all the WNs are non-degenerate.
Interestingly, the position of WNs 
in BZ changes along with their energies by changing the orientation of MQA while the total number of WNs in the system remains unaltered \cite{sm_text} (see Figs. \ref{fig:stuc_wp} (b), (c)). However, inside a given energy window around the Fermi energy, the number of  WNs with positive and negative  chiralities do not match due to the lifting of degeneracy. As a result, a chirality imbalance is established within a limited energy window. \textcolor{black}{Note that} there exist no chirality imbalance in the entire energy landscape due to the presence of equal numbers of  WNs with opposite chiralities. We below focus on the effect of the degeneracy lifting on the magneto-transport in PrRhC$_2$ considering the PHE steup. 
\textcolor{black}{In particular, we compute velocity and Berry curvature from tight-binding Wannier Hamiltonian, corresponding to a given MQA, that we further use for the calculation of PH transport \cite{sm_text}.} We interchangeably use MQA $[100]$, $[011]$ and $[111]$ for  $\bf{q}$-$100$, $\bf{q}$-$011$ and $\bf{q}$-$111$, respectively.

\begin{table}[b!]
\caption{Magnetic symmetry group with the degeneracy of the WNs for different MQA. }
\begin{tabular*}{0.45\textwidth}{ p{2.0cm} p{4.1cm} p{3.5cm}} 
    \hline
    \hline
        {$\bf{q}$-axis} &  Mag. Grp. Symm. & Degeneracy\\
    \hline
    \hline
    non-magnetic & $\{E$, $m(x)$, $m(y)$,$C_2(z)$, $\mathcal{T}$,    &  \\
        case  &  $m(x)\mathcal{T}$, $m(y)\mathcal{T},C_2(z)\mathcal{T}\}$   &~~~ 8 \\
        1 0 0 & $\{E$, $m(x)$, $m(y)\,\mathcal{T}, C_2(z)\,\mathcal{T}\}$   &~~~ 4 \\
        0 1 0 & $\{E$, $m(x)\,\mathcal{T}$, $m(y),C_2(z)\,\mathcal{T}\}$   &~~~ 4 \\
        0 0 1 & $\{E$, $m(x)\,\mathcal{T}$, $m(y)\,\mathcal{T},C_2(z)\}$   &~~~ 4 \\
        0 1 1 & $\{E$, $m(x)\,\mathcal{T}\}$     &~~~ 2  \\
        1 0 1 & $\{E$, $m(y)\,\mathcal{T}\}$     &~~~ 2  \\
        1 1 0 & $\{E$, $C_2(z)\,\mathcal{T}\}$     &~~~ 2  \\
        1 1 1 & $\{E\}$  &~~~   1  \\
    \hline
\end{tabular*}
\label{tab:aniso}
\end{table}

{\it \blue {{{Planar Hall effect}} for B$\neq$0:}} 
PrRhC$_2$ constitutes  alternative layers of Pr and RhC$_2$ along the $x$-axis.  Each layer forms an approximate triangular lattice in $yz$-plane  \cite{sadhukhan}. In order to study the PHE, we consider the electric field $\mathbf{E}$ along the $y$-axis and magnetic field $\mathbf{B}$ lying in $yz$-plane at a finite angle $\gamma$ with respect to the $y$-axis:  $\mathbf{E}=E\hat{y}$, $\mathbf{B}=B\cos\gamma \hat{y}+B\sin\gamma\hat{z}$. 
Following the semi-classical Boltzmann transport equation and relaxation time approximation, the PHC $\sigma_{zy}$ and LMC $\sigma_{yy}$ are found to be   
\cite{Nandy_2017, Nandy_2018, Tanay_2018, Sharma:2016}
\ba
\sigma_{zy}&\simeq e^{2}& \displaystyle \int\frac{d^{3}k}{(2\pi)^{3}}D\tau\left(-\frac{\partial f_{0}}{\partial \epsilon}\right) 
\Big[ \big(v_{z}+\frac{eB\sin \gamma}{\hbar}(\mathbf{\Omega_{k}}\cdot\mathbf{v_{k}})\big) \nonumber \\
&\times &\big(v_{y}+\frac{eB\cos \gamma}{\hbar}(\mathbf{\Omega_{k}}\cdot\mathbf{v_{k}})\big)\Big]
\label{eq_ehc}
\ea \textcolor{black}{and} 
\begin{eqnarray}
\sigma_{yy}&& \simeq e^{2} \displaystyle \int\frac{d^{3}k}{(2\pi)^{3}}D \tau \left(-\frac{\partial f_{0}}{\partial \epsilon}\right) \Big({v_{y}}+
\frac{eB\cos \gamma}{\hbar}(\mathbf{\Omega_{k}}\cdot\mathbf{v_{k}})\Big)^{2}  \nonumber \\
\label{eq_lmc}
\end{eqnarray}
where $D\equiv D(\mathbf{B,\Omega_{k}})=(1+\frac{e}{\hbar}(\mathbf{B}\cdot \mathbf{\Omega_{k}}))^{-1}$ is the phase space
factor~\cite{Duval_2006}. The Berry curvature and velocity are denoted by $\mathbf{\Omega_{k}}=(\Omega_x,\Omega_y,\Omega_z)$ and  $\mathbf{v_{k}}=(v_x,v_y,v_z)$, respectively.

Notice that the chiral magnetic effect can lead to finite charge current for WSM in the absence of $\mathbf{E}$. The factor $(\mathbf{\Omega_{k}}\cdot\mathbf{v_{k}})\mathbf{B}$ is found to be responsible for chiral magnetic effect. On the other hand, in the presence of $\mathbf{E}$, the charge current becomes proportional to the factor $\mathbf{E}\cdot\mathbf{B}$ which is primarily responsible for chiral anomaly. The left (right) WN is over (under) populated and a current flows obeying the direction of $\mathbf{E}$ (see Fig. \ref{fig:phe_temp}(a)).
Importantly, $eB\sin \gamma$ and $eB\cos \gamma$ factors are originated from anomalous velocity $(\mathbf{E} \times \mathbf{\Omega_{k}}) \cdot\mathbf{B}$ and chiral anomaly $\mathbf{E}\cdot\mathbf{B}$ contributions, respectively. In the present case,
the chirality imbalance, caused by  degeneracy lifting of  WNs, around the Fermi surface plays crucial role through the  factor $\frac{\partial f_{0}}{\partial \epsilon}$ in PHC and LMC.


In order to work with the Boltzmann \textcolor{black}{transport} formulation, we consider temperature $ T \sim  10^{-3}$ eV, magnetic field $B\sim  10^{-5}$ eV  and Fermi energy $\mu \sim 10^{-2}$ eV satisfying  $T \ll \sqrt{B} \ll \mu$. We choose $k$-grids  $300 \times 300 \times 300$ for our numerical calculations where we obtain a satisfactory convergence within $\sim$ (3-5)\% \cite{sm_text}.  Note that  ${D}=(1+\frac{e}{\hbar}(\mathbf{B}\cdot \mathbf{\Omega_{k}}))^{-1}$ remains unity all over the BZ except at the WNs where Berry curvature becomes substantially large; we consider $D=1$ throughout for the sake of numerical convergence. \textcolor{black}{One can carefully analyse the $\mathbf{B}$-dependence considering $D\ne 1$ in future.} {\mg{The relaxation time $\tau$ is of the order of femtosecond, predicted for the metallic systems, and is also applicable for the present case  \cite{PhysRevB.94.155105}. The anisotropic nature of the Fermi surface, as shown in Fig.~\ref{fig:stuc_wp} (b), is expected to result in anisotropy in relaxation time that we neglect for our magnetotransport calculations for simplicity. }} We compute the conductivities after subtracting the zero field part:  $(\sigma_{ij}(B)-\sigma_{ij}(B=0))/\sigma_{ij}(B=0)$, where $i,j=x,y,z$.

Figure \ref{fig:phe_temp} (b) shows the variation of PHC with temperature for different ${\bf q}$'s where we find highly non-linear growth  with $T$ for $T\gtrsim30$ K.
We, therefore, restricts ourselves between  $22~ {\rm K}<T < 26 ~{\rm K}$ where  PHC acquires finite value
with $T^2$-dependence as predicted by the Sommerfeld expansion \cite{Tanay_2018} (see inset of Fig. \ref{fig:phe_temp} (c)).
The PHCs for ${\bf q}$-$011$ and -$111$ are of the order of (m$\Omega$.cm)$^{-1}$ but for the case of ${\bf q}$-$100$, it is of the order of ($\mu \Omega$.cm)$^{-1}$. This can be understood by the fact that the degeneracy of the WNs is lifted upon changing the orientation of MQA from easy axis to body-diagonal.
We explicitly calculate the contributions from the individual terms in PHC (Eq.~(\ref{eq_ehc})) for MQA $[111]$ as shown in Fig. \ref{fig:phe_temp} (c) \cite{sm_text}.  We note that the term comprising of  chiral magnetic effect factor $(\mathbf{\Omega_{k}}\cdot\mathbf{v_{k}})^2$ acquires maximum value as compared to the remaining terms. This in turn refers to the fact that the term $B^2 \sin\gamma \cos \gamma (\mathbf{\Omega_{k}}\cdot\mathbf{v_{k}})^2 $, associated with the chiral anomaly, dictates the topological Hall response in presence of electric and magnetic fields.
We, therefore, demonstrate the momentum resolved structure of  $(\Omega \cdot v) \equiv \sum \int dk_x(\mathbf{\Omega_{k}}\cdot\mathbf{v_{k}}) $ ($\sum$ includes the filled band only) to highlight its distribution over the ($k_y-k_z$)-plane (see Fig. \ref{fig:phe_temp}(d)) \cite{sm_text}.

We now co\textcolor{black}nsider a  moderate temperature $T=25$ K to investigate the PHE in terms of the angular and magnetic field dependence as shown in  Figs. \ref{fig:phe} (a), (c) and (b), (d), respectively.
The LMC has the angular  dependence   $\cos^2 \gamma$ while the PHC follows nearly $\sin\gamma \cos\gamma$-dependence. On the other hand, $B^2$ \textcolor{black}{profile} is commonly observed for
$\sigma_{yy}$ and $\sigma_{zy}$. \textcolor{black}{The type-II WSMs can exhibit linear variation of PHC and LMC with magnetic field \cite{Nandy_2017,Jiang19,Wang_2018} in addition to the quadratic one that is found to be predominant in our case. Our findings on the angular signature of PHE are in accordance with the experimental observations \cite{Wang_2018, Kumar_2018}. } Most importantly,  the LMC and PHC exhibit strongest (weakest) response when the internal MQA aligns with the  body diagonal (easy axis). Notice that the non-degenerate (maximally degenerate) WNs are only observed for ${\bf q}$-$111$ (${\bf q}$-$100$) case. An intermediate response is noticed for the case ${\bf q}$-$011$ where WNs are $2$-fold degenerate.

Having discussed the numerical findings, we below understand the chiral anomaly contributions with plausible arguments.
Considering the generic low-energy  dispersion of WSM $\epsilon=-k+ \xi k_z =k(-1+\xi \cos\theta)$ (with $\xi$ being the tilt parameter along $z$-direction and $k= \sqrt{k_x^2+k_y^2+k_z^2}$)  around an isolated WN of topological charge $C$, one can analytically compute the quantity $A=\int d^{3}k (\mathbf{v_{k}}\cdot\mathbf{\Omega_{k}})^2 \frac{\partial f_{0}}{\partial \epsilon}$ associated with dominant chiral anomaly term in the PHC and LMC. We find $ A \approx  C(2\mu^{-2} + 2 \xi^2 \mu^{-2} +{\mathcal O}(T^2)  )$.  We consider $D=1$, $\frac{\partial f_{0}}{\partial \epsilon}\approx \delta (\theta-\theta')/|\xi k \sin\theta| +{\mathcal O}(T^2)$  and $d^3k=k^2 \sin\theta d\theta d\phi dk$ while deriving the above approximate form in the limit $B\ll \mu^2$ \cite{Tanay_2018}. The tilt $\xi$ shrinks the limits of $k$-integration. 
In presence of $m$ number of  non-degenerate WNs of chiralities $C_m$ appearing at  energies $\mu_m$, the quantity $A$ can be considered in the following form: 
$A\approx \sum_m 2 C_m (\mu^{-2}_m +\xi^2 \mu^{-2}_m +{\mathcal O}(T^2))$. 
Once the degeneracy of the WNs is lifted, the  WNs with positive and negative chirality no longer reside at the same energy. This corresponds to $C_i =-C_j$ and $\mu_i \ne \mu_j$ ensuring $A\propto (\mu_i-\mu_j)$. One can thus find that the contributions from $(\mathbf{v_{k}}\cdot\mathbf{\Omega_{k}})^2$ terms in PHC and LMC enhance with the degeneracy lifting as originated from the change in the orientation of MQA. As a result,  
$B^2 \cos^2\gamma$ ($B^2 \sin\gamma \cos \gamma)$ factor, associated with $A$, is found to  play the  governing role in the LMC (PHC). \textcolor{black}{Noticeably, $A$ itself increases when MQA is tuned from  $[100]$ to $[111]$}. This reminds the behavior of charge current due to chiral magnetic effect that is found to be $J= (e^2/\hbar) \mathbf{B} \int d^{3}k (\mathbf{v_{k}}\cdot\mathbf{\Omega_{k}}) f_{0} 
\approx -(e^2/\hbar^2) \mathbf{B}\sum_{m}\mu_{m}C_{m}$ for the un-tilted WSM at $T=0$ K \cite{PhysRevB.88.104412,PhysRevLett.116.077201,Son_2012,Zhou_2013,PhysRevB.92.085138,sadhukhan}.

The chiral anomaly induced charge current of the order of ($\mu\Omega$.cm)$^{-1}$ is observed for PHE setup with MQA $[100]$. What is more interesting in the present case is that charge current can be amplified to  (m$\Omega$.cm)$^{-1}$ by varying MQA from high-symmetry to low-symmetry direction. The degeneracy lifting induces the chiral chemical potential $\mu_{\rm ch}=\mu_+ -\mu_-$ between WNs of topological charge $C_{\pm}$ and  $\mu_{\rm ch}$ effectively renormalizes the external electric field $\mathbf {E} \to \mathbf{E} + \nabla \mu_{\rm ch}$. This refers to the increment in $ \mathbf{B} \cdot \mathbf{E} $ factor.  Therefore, the chiral anomaly (on-field) and chirality imbalance (off-field) both  imprint their effects on  PHC and LMC while changing the MQA appropriately.

{\it \blue{{{Anomalous Hall effect}} for $B=0$:}} 
We shall now investigate the AHC $\sigma^a_{ij}$ given by 
\be
\sigma^a_{ij}= -\frac{e^2}{\hbar} \sum_{n=1}^{\rm occupied} \int d^3 k ~\Omega^{n,k}_{\mathbf{k}} f_0 (\epsilon_n)
\ee
as a function of chemical potential $\mu$ where $i,j,k=x,y,z$.
We show their behavior in Fig. \ref{fig:ahe} (a), (b), and (c)
for the MQA along $[100]$, $[011]$ and $[111]$, respectively. The important point to notice is that $\sigma^a_{yz}$
becomes finite for $[100]$ and $[111]$  \textcolor{black}{and} vanishes for $[011]$. Interestingly, $\sigma^a_{xy}$ and $\sigma^a_{xz}$  become vanishingly small for  $[100]$
while finite for  $[011]$ and $[111]$.
The most pronounced responses are observed for 
$\sigma^a_{yz} \approx 94$, $\sigma^a_{xz} \approx 87$, and $\sigma^a_{xy} \approx -70$ (units of $(\Omega {\rm cm})^{-1}$) in $[100]$, $[011]$ and $[111]$ around $\mu \approx 0.18$ eV,  respectively. 
Another noticeable feature is that $\sigma^a_{xy}$ and $\sigma^a_{xz}$ nearly appear in opposite sign for $[011]$ and $[111]$.  The AHC does not
\textcolor{black}{acquire} high value around the WNs energies referring to the fact that the $\mathbf{k}$-space separation between WNs play important role in determining AHC than the energies of WNs. This is in congruence with the results on  
AHC  as obtained from the low-energy model of WSM \cite{PhysRevB.86.115133,Burkov:2011}. {\mg{Note that we only compute the intrinsic part of the AHC and neglect the extrinsic parts namely, skew scattering and side-jump as caused by impurity scattering \cite{RevModPhys.82.1539}. Since we consider a clean system, the magnitudes of the above extrinsic quantities might be significantly less as compared to the intrinsic part.}}

We focus on the symmetry aspects, associated with the substantially large components of AHC, \textcolor{black}{by} lifting the degeneracy  of the WNs by changing MQA.  For  ${\bf q}$-$100$ as shown in Fig. \ref{fig:ahe} (a),  $\sigma^a_{yz}$ [$\sigma^a_{xy}$, and $\sigma^a_{xz}$] acquires [acquire] high [low] value allowed by  $C_2(z)\,\mathcal{T}$, $m(x)$, $m(y)\,\mathcal{T}$ symmetries. One can find that under $C_2(z)$, $\Omega_z(k_x,k_y,k_z) \to \Omega_z{(-k_x,-k_y,k_z)}$ and $\Omega_{x,y}(k_x,k_y,k_z) \to  -\Omega_{x,y}(-k_x,-k_y,k_z)$. For  $C_2(z)\,\mathcal{T}$ symmetry, $\Omega_z(k_x,k_y,k_z) \to -\Omega_z{(k_x,k_y,-k_z)}$ and 
$\Omega_{x,y}(k_x,k_y,k_z) \to 
\Omega_{x,y}{(k_x,k_y,-k_z)}$. This results in the vanishingly small response for $\sigma^a_{xy}$ while summing 
$\Omega_z$ over the BZ. Next, for $m(y)$,  
$\Omega_y(k_x,k_y,k_z) \to  \Omega_y(k_x,-k_y,k_z)$ and 
$\Omega_{x,z}(k_x,k_y,k_z) \to - \Omega_{x,z}(k_x,-k_y,k_z)$. Under  $m(y)\,\mathcal{T}$ symmetry, $\Omega_y(k_x,k_y,k_z) \to  -\Omega_y(-k_x,k_y,-k_z)$ and 
$\Omega_{x,z}(k_x,k_y,k_z) \to \Omega_{x,z}(-k_x,k_y,-k_z)$. This
further leads to vanishingly small  value of $\sigma^a_{xz}$. Last, for $m(x)$, $\Omega_x(k_x,k_y,k_z) \to \Omega_x(-k_x,k_y,k_z)$ and 
$\Omega_{y,z}(k_x,k_y,k_z) \to -\Omega_{y,z}(-k_x,k_y,k_z)$. This causes
a substantial contribution in 
$\sigma^a_{yz}$.

For $\bf{q}$-$011$,  the symmetry  $m(x)\,\mathcal{T}$ allows 
$\Omega_x(k_x,k_y,k_z) \to  -\Omega_x(k_x,-k_y,-k_z)$ and 
$\Omega_{y,z}(k_x,k_y,k_z) \to \Omega_{y,z}(k_x,-k_y,-k_z)$. Therefore, $\sigma^a_{yz}$ is thus constrained to vanish \textcolor{black}{and} the remaining AHCs continue acquiring finite  values. For ${\bf q}$-$111$, there no longer exists any symmetry in the system. \textcolor{black}{As a result}, all components of the Berry curvature contribute yielding $\sigma_{xy}$, $\sigma_{yz}$, and $\sigma_{xz}$ to be non-zero.

We further note that 
chirality imbalance can determine the profile of AHC as the distribution of Berry curvature in BZ changes with MQA. 
It is evident from AHCs $\sigma^a_{xz}$ and $\sigma^a_{xy}$ that $\Omega_{y}$ and $\Omega_{z}$, summed over the filled bands, become maximally negative and positive in the BZ, respectively, for  $\bf{q}$-$011$ (see Fig. \ref{fig:ahe} (b)).
The net sign of the band-summed $\Omega_{x,y,z}$ in BZ might be related to the net chirality of the WNs below the highest filled bands. 
However, notice that the chirality of a WN is determined by all three components of the Berry curvature not by an individual component.
For $\sigma^a_{yz}$, positive and negative regions of $\Omega_x$ over the BZ cancel each other that result in an infinitesimally small response under $\bf{q}$-$011$ when summing over the filled bands. Interestingly, positive and negative regions of $\Omega_x$ do not cancel each other in the BZ for the filled bands leading to positive and negative values of $\sigma^a_{yz}$ under $\bf{q}$-$100$ (see Fig. \ref{fig:ahe} (a)).   
The imbalance in the number of opposite chirality WNs can play crucial role in the anomalous transport
within a given window of chemical potential
in absence of any external magnetic and electric fields.


{\it \blue {Summary:}} 
\textcolor{black}{The imbalance between the number of opposite chirality WNs around the  Fermi energy is engineered by changing the MQA in non-centrosymmetric WSMs family  \cite{sadhukhan}.
This motivates us to}
probe the interplay between topology and magnetism in the transport properties for a candidate material PrRhC$_2$ within the above family of materials.  We find that the underlying 
degeneracy of WNs can be systematically lifted 
in accordance with the symmetries upon 
\textcolor{black}{tuning} the MQA from $[100]$ to $[111]$ (see Fig. \ref{fig:stuc_wp}). This leads to a mismatch  between the number of positive and negative chirality WNs near the Fermi energy that is 
referred to as the chirality imbalance. Under the application of co-planar electric and magnetic fields in PH setup,  we find the hallmark angular  dependence in PHC and LMC suggesting to chiral anomaly induced topological transport in the material (see Figs. \ref{fig:phe_temp} and \ref{fig:phe}). \textcolor{black}{The above finding is consistent with experiment \cite{Wang_2018, Kumar_2018}, however, the associated quadratic dependence on magnetic field is a subject of further investigation.}  Interestingly, the degeneracy lifting causes this transport to become more pronounced due to the combined effect of chirality imbalance and chiral anomaly (see  Fig. \ref{fig:phe}). Last, we find symmetry permitted coefficients of AHC under different MQA where the sign of the AHCs might be related to the chirality imbalance within a limited energy window (see
Fig. \ref{fig:ahe}). Notice that AHC is substantially large [units of  ($\Omega$.cm)$^{-1}$] as compared to the PHC [units of  ($\mu\Omega$.cm)$^{-1}$ and (m$\Omega$.cm)$^{-1}$] suggesting the validity of our results from the experimental perspective \cite{PhysRevB.98.121108,  PhysRevB.98.041103, Taskin2017, PhysRevB.97.201110}. Our study can further  stimulate
transport experiments specially in rare-earth transition metal carbides RMC$_2$ (R = rare earth and M = transition metal) family to probe the chirality imbalance by PHE.
The magneto-transport can be investigated in the  quantum limit using the framework of DFT in future \cite{xiong2022understanding}. {\mg{Motivated by the studies on graphene \cite{RevModPhys.83.407}, one can analyze 
the chiral anomaly induced magnetoconductivity, thermal Hall conductivity and Wiedemann-Franz law by considering the more realistic
momentum-dependent relaxation time in future. }}

{\it \blue {Acknowledgments:}} 
BS thanks Rajyavardhan Ray for providing DFT structures and Ulrike Nitzsche for technical assistance. We thank Jeroen van den Brink, Manuel Richter, and Jorge I. Facio for discussion during the initial stage of the project. We acknowledge  cluster facility provided by IFW-ITF, Dresden.

\bibliography{phe}

\end{document}